# Transfer of Orbital Angular Momentum in Vortex Light through Four-Wave Mixing and the Manipulation of Slow and Fast Light

Fan Meng, Xin-Yao Huang[†], Guo-Feng Zhang[*]

*School of Physics, Beihang University, Beijing 102206, China*

**Abstract**

Vortex light, a unique optical field that carries orbital angular momentum (OAM), has attracted considerable attention in recent years. In this paper, we present a detailed theoretical analysis of OAM transfer from the input field to the generated signal field in a four-level double-Λ system via the four-wave mixing (FWM) process, showing that their OAMs follow a specific algebraic relationship. We identify the optimal conditions for efficient vortex light transmission, analyze the influence of detuning on transmission efficiency and phase distortion, and specifically examine the scenario where the control field $\Omega_{c2}$ carries OAM—the latter being essential for a complete characterization of OAM conservation in the FWM process, while all three aspects have been largely overlooked in the existing literature. Furthermore, we investigated the tunability of the group velocity between the probe and signal fields by modulating the Rabi frequencies of the two control fields and the relative phase between the probe and signal fields during the FWM process. We demonstrate that the conversion between matched vortex slow and fast light can be realized—an effect that has not been widely explored in dual-Λ-type systems. These results may hold promise for applications in quantum information storage and processing, quantum computing, and ultrasensitive detection.

## 1. Introduction

Researchers are actively investigating the optical properties of quantum systems through quantum interference effects [1-3]. One of the key phenomena is electromagnetically induced transparency (EIT), a quantum coherence effect that enables light pulses to traverse an otherwise opaque medium [4-7]. EIT has been extensively investigated both theoretically and experimentally, yielding significant applications in various domains [8-13]. A notable phenomenon associated with EIT is its transformative impact on the absorption and dispersion characteristics of the medium, leading to normal or anomalous dispersion at resonance. These dispersion regimes correspond to a deceleration ($0 < v_g \ll c$) or acceleration ($v_g > c$ or $v_g < 0$) of the light's group velocity, respectively [14,15]. Significant progress has been made in the exploration of fast and slow light in dispersive media, with numerous studies delving deeply into this intriguing field [16-18]. Naturally, the regulation of fast and slow light extends beyond EIT, as alternative approaches have also been proposed [19,20].

On the one hand, Harris et al. were the first to propose utilizing the EIT effect to achieve slow light [21]. Experimentally, Hau et al. successfully reduced the group velocity of light to 17 m/s in a Bose-Einstein condensate [22]. Subsequently, in a hot Rb atomic vapor, the group velocity was further reduced to 90 m/s [23] and later to 8 m/s [24]. The group velocity of light reached as low as 45 m/s in a Pr-doped $Y_2SiO_5$ optically dense crystal [25]. Research on slow light extends beyond mere group velocity reduction. It also serves as a crucial mechanism for storing and freezing light within a medium, enabling reversible quantum memory, which is expected to play a pivotal role in future quantum information technologies. Lukin et al. developed the dark polaron theory in media and proposed the feasibility of achieving optical storage in a medium via the EIT effect [26]. Utilizing the slow light effect, Beil et al. demonstrated the immense potential of feedback-controlled pulse shaping for optical storage in EIT-driven solid-state media [27]. The deceleration and storage of compressed vacuum pulses were successfully realized through EIT in cold atomic systems [28]. Recently, an alternative approach distinct from EIT has been proposed, enabling the simultaneous capture of two light pulses in atomic ensembles via the static light pulse (SLP) process [29].

---

[†] Corresponding author. E-mail: xinyaohuang@buaa.edu.cn
[*] Corresponding author. E-mail: gf1978zhang@buaa.edu.cn

On the other hand, the propagation of light at superluminal speeds has garnered significant attention. Chu et al. were the first to observe the superluminal phenomenon in a resonant system composed of CaP:N samples, where the pulses propagated through the medium without shape distortion but underwent strong resonant absorption [30]. Theoretically, superluminal propagation of laser pulses between two gain lines can be observed in an inverted medium with a two-line structure [31]. Experimental studies on superluminal propagation using gain-assisted linear anomalous dispersion in Cs atomic gas demonstrate this phenomenon [32]. In this case, the group velocity of laser pulses not only surpasses the speed of light $c$ but can even attain negative values. The anomalous propagation of electromagnetic wave packets traveling at negative group velocities within plasma frequency bands, far from the absorption resonance zone, has been experimentally observed [33]. Regarding superluminal light, it is widely accepted, following the view of Sommerfeld and Brillouin et al. [34-36], that although the pulse envelope can be adjusted to propagate in the medium at a certain group velocity—making the pulse peak appear to move faster than the speed of light $c$—the transmission of information, as a fundamental constraint, remains limited to $c$ or lower. Consequently, the superluminal propagation of the pulse peak does not violate Einstein's special theory of relativity. It is argued that true information is carried by the wavefront, which does not exceed the speed of light $c$ in vacuum. This perspective has been further refined, suggesting that true information is encoded at the non-analytic points of wave packets [37-39]. The signal front can be regarded as one such non-analytic point. Fast light holds potential for applications in all-optical information processing [40,41] and ultra-sensitive measurements [17].

As a structured light field with a distinctive phase structure, vortex light carries a helical phase factor $e^{il\varphi}$, where $l$ is known as the topological charge and can take integer or non-integer values, while $\varphi$ denotes the azimuthal angle in the beam's cross-section. Such a beam carries an orbital angular momentum (OAM) of $l\hbar$ per photon, leading to a variety of intriguing and significant phenomena when interacting with a medium [42-59]. Vortex light finds significant applications not only in quantum information science, particularly in quantum storage, transmission, and processing [43,47,48,50-52], but also in diverse fields such as optical tweezers [60], optical testing [61], optical processing [62], and microscopy imaging [63]. The potential of this field can be further expanded by incorporating vortex light into the study of fast and slow light phenomena [64,65]. The introduction of OAM introduces an additional degree of freedom for manipulating and harnessing fast-slow light, making this phenomenon particularly intriguing as it enables precise control over optical flow. Most prior studies have primarily investigated the transfer of vortex light's OAM across different channels in light-matter interactions, whereas relatively few have explored the regulation of fast and slow vortex light via dispersion of the medium. Recent theoretical studies have discussed the propagation of vortex slow light in tripod-coherent preparation media [66]. However, these works primarily focus on slow-light propagation and lack a analysis of fast-slow light regulation during vortex light conversion. In this paper, we employ a four-level dual-Λ system to investigate the transfer of OAM during four-wave mixing (FWM) and the associated variations in fast and slow light. Our research differs from previous studies in several key aspects: (1) We investigate the scenario in which all three input fields are vortex beams propagating through the medium, and demonstrate that the previously overlooked control field, $\Omega_{c2}$, can also carry OAM; (2) we identify the physical conditions required for efficient vortex optical transmission; (3) we analyze the influence of detuning on transmission efficiency and phase distortion during the OAM transfer process; and (4) we demonstrate the switching between matched vortex slow and fast light by tuning relevant parameters within the FWM process. These studies further enrich our understanding of vortex light-matter interactions and may have implications for quantum computing, as well as quantum information storage and transmission.

## 2. Theory and Analysis

### A. Dynamical Behavior of the System

As illustrated in Fig. 1, we analyze a closed quantum system exhibiting a four-level dual-Λ configuration. In this setup, the strong fields $\Omega_{c1}$ and $\Omega_{c2}$ drive the atomic transitions $|2\rangle \leftrightarrow |4\rangle$ and $|2\rangle \leftrightarrow |3\rangle$, respectively, whereas the weak probe field $\Omega_p$ and the generated FWM signal field $\Omega_s$ couple to the transitions $|1\rangle \leftrightarrow |3\rangle$ and $|1\rangle \leftrightarrow |4\rangle$, respectively. The decay rates of the excited state $|3\rangle$ to the ground states $|1\rangle$ and $|2\rangle$ are denoted as $\Gamma_{31}$ and $\Gamma_{32}$, respectively, while the decay rates of the excited state $|4\rangle$ to the ground states $|1\rangle$ and $|2\rangle$ are given by $\Gamma_{41}$ and $\Gamma_{42}$, respectively. $\Gamma_{12}$ denotes the relaxation rate between the ground states $|1\rangle$ and $|2\rangle$. By applying the rotating-wave approximation and the dipole approximation, the system Hamiltonian can be formulated as follows:

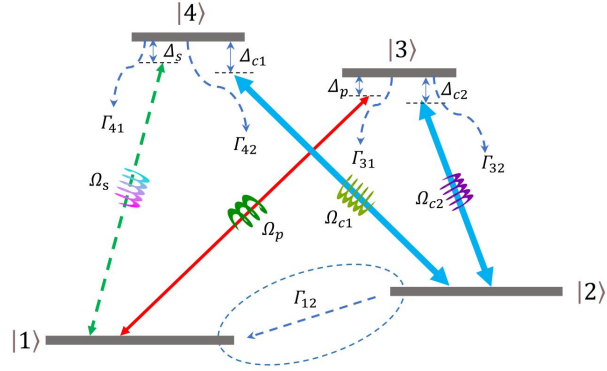

**Fig. 1.** Schematic representation of a four-level system with a dual-Λ configuration.

$$H = -\hbar(\Delta_p - \Delta_{c2})|2\rangle\langle 2| - \hbar\Delta_p|3\rangle\langle 3| - \hbar(\Delta_p - \Delta_{c2} + \Delta_{c1})|4\rangle\langle 4| \\ -\hbar(\Omega_p e^{i\mathbf{k}_p \cdot \mathbf{r}}|3\rangle\langle 1| + \Omega_{c2} e^{i\mathbf{k}_{c2} \cdot \mathbf{r}}|3\rangle\langle 2| + \Omega_s e^{i\mathbf{k}_s \cdot \mathbf{r}}|4\rangle\langle 1| + \Omega_{c1} e^{i\mathbf{k}_{c1} \cdot \mathbf{r}}|4\rangle\langle 2| + \text{H.c.}) \quad (1)$$

where $\Delta_p = \omega_p - \omega_{31}$, $\Delta_{c2} = \omega_{c2} - \omega_{32}$ and $\Delta_{c1} = \omega_{c1} - \omega_{42}$ denote the respective detuning parameters. The parameters $\omega_p$, $\omega_{c1}$ and $\omega_{c2}$ correspond to the central frequencies of the three input optical fields, while $\omega_{31}$, $\omega_{32}$ and $\omega_{42}$ represent the transition frequencies between the respective atomic energy levels. $\Omega_p = \mu_{31}E_p/2\hbar$, $\Omega_{c2} = \mu_{32}E_{c2}/2\hbar$, $\Omega_s = \mu_{41}E_s/2\hbar$ and $\Omega_{c1} = \mu_{42}E_{c1}/2\hbar$ denote the respective Rabi frequencies of the corresponding optical fields. $\mu_{mn}(m=3,4;n=1,2)$ denotes the electric dipole matrix element, and $E_j(j=p,c1,c2,s)$ represents the amplitude of the optical field. The position and wave vectors of the corresponding optical field are denoted by $\mathbf{r}$ and $\mathbf{k}_j(j=p,c1,c2,s)$, respectively. The system Hamiltonian is formulated within the framework of the rotating optical frequency, imposing a constraint on the detuning parameters of the four interacting fields, given by $\Delta_s = \Delta_p - \Delta_{c2} + \Delta_{c1}$. We assume that the two control fields are resonant, i.e., $\Delta_{c1} = \Delta_{c2} = 0$.

The dynamical evolution of a four-level dual-$\Lambda$ system can be characterized by Liouville equation as follows:

$$\frac{\partial}{\partial t}\rho = L_\rho - \frac{i}{\hbar}[H,\rho], \qquad (2)$$

here the first term on the right-hand side of the equation accounts for the damping effect, while the second term describes the primary dynamical process.

By substituting Eq. (1) into Eq. (2), we derive the explicit equation of motion for this four-level system:

$$\frac{\partial}{\partial t}\rho_{11} = \Gamma_{31}\rho_{33} + \Gamma_{41}\rho_{44} + \Gamma_{12}\rho_{22} + i\Omega_p^*\rho_{31} + i\Omega_s^*\rho_{41} - i\Omega_p\rho_{13} - i\Omega_s\rho_{14},$$

$$\frac{\partial}{\partial t}\rho_{22} = \Gamma_{32}\rho_{33} + \Gamma_{42}\rho_{44} - \Gamma_{12}\rho_{22} + i\Omega_{c2}^*\rho_{32} + i\Omega_{c1}^*\rho_{42} - i\Omega_{c2}\rho_{23} - i\Omega_{c1}\rho_{24},$$

$$\frac{\partial}{\partial t}\rho_{33} = -\Gamma_{31}\rho_{33} - \Gamma_{32}\rho_{33} + i\Omega_p\rho_{13} + i\Omega_{c2}\rho_{23} - i\Omega_p^*\rho_{31} - i\Omega_{c2}^*\rho_{32},$$

$$\frac{\partial}{\partial t}\rho_{44} = -\Gamma_{41}\rho_{44} - \Gamma_{42}\rho_{44} + i\Omega_s\rho_{14} + i\Omega_{c1}\rho_{24} - i\Omega_s^*\rho_{41} - i\Omega_{c1}^*\rho_{42},$$

$$\frac{\partial}{\partial t}\rho_{21} = -\frac{\Gamma_{12}}{2}\rho_{21} + i(\Delta_p - \Delta_{c2})\rho_{21} + i\Omega_{c2}^*\rho_{31} + i\Omega_{c1}^*\rho_{41} - i\Omega_p\rho_{23} - i\Omega_s\rho_{24},$$

$$\frac{\partial}{\partial t}\rho_{31} = -\frac{\Gamma_{31}+\Gamma_{32}}{2}\rho_{31} + i\Delta_p\rho_{31} + i\Omega_p(\rho_{11}-\rho_{33}) + i\Omega_{c2}\rho_{21} - i\Omega_s\rho_{34},$$

$$\frac{\partial}{\partial t}\rho_{41} = -\frac{\Gamma_{41}+\Gamma_{42}}{2}\rho_{41} + i(\Delta_p - \Delta_{c2} + \Delta_{c1})\rho_{41} + i\Omega_s(\rho_{11}-\rho_{44}) + i\Omega_{c1}\rho_{21} - i\Omega_p\rho_{43},$$

$$\frac{\partial}{\partial t}\rho_{32} = -\frac{\Gamma_{31}+\Gamma_{32}+\Gamma_{12}}{2}\rho_{32} + i\Delta_{c2}\rho_{32} + i\Omega_{c2}(\rho_{22}-\rho_{33}) + i\Omega_p\rho_{12} - i\Omega_{c1}\rho_{34},$$

$$\frac{\partial}{\partial t}\rho_{42} = -\frac{\Gamma_{41}+\Gamma_{42}+\Gamma_{12}}{2}\rho_{42} + i\Delta_{c1}\rho_{42} + i\Omega_{c1}(\rho_{22}-\rho_{44}) + i\Omega_s\rho_{12} - i\Omega_{c2}\rho_{43},$$

$$\frac{\partial}{\partial t}\rho_{43} = -\frac{\Gamma_{31}+\Gamma_{32}+\Gamma_{41}+\Gamma_{42}}{2}\rho_{43} + i(\Delta_{c1}-\Delta_{c2})\rho_{43} + i\Omega_s\rho_{13} + i\Omega_{c1}\rho_{23} - i\Omega_p^*\rho_{41} - i\Omega_{c2}^*\rho_{42}$$

$$(3)$$

As we consider a closed system, the conservation of particle number must be satisfied: $\rho_{11} + \rho_{22} + \rho_{33} + \rho_{44} = 1$. Additionally, the Hermitian property must hold:

$\rho_{ij} = \rho_{ji}^*$. For simplicity, we assume $\Gamma_{12} = 0$, $\Gamma_{31} = \Gamma_{32} = \Gamma_3/2$, $\Gamma_{41} = \Gamma_{42} = \Gamma_4/2$, and set $\Gamma_3 = \Gamma_4 = \Gamma$.

## B. Transfer of OAM in Vortex Light

In the chosen four-level system, we assume that the probe field $\Omega_p$ and the signal field $\Omega_s$ are significantly weaker than the two control fields $\Omega_{c1}$ and $\Omega_{c2}$, i.e., $\Omega_p, \Omega_s \ll \Omega_{c1}, \Omega_{c2}$. Under this assumption, the control fields can be regarded as always present and remaining constant with respect to time $t$ and spatial coordinate $z$. Additionally, we assume that the system is initially in the ground state $|1\rangle$, specifically $\rho_{11}^{(0)} = 1$, $\rho_{22}^{(0)} = \rho_{33}^{(0)} = \rho_{44}^{(0)} = 0$ and $\rho_{ij}^{(0)} = 0$ $(i \neq j)$. By applying a perturbative approximation ($\rho_{ij} = \rho_{ij}^{(0)} + \Omega_p \rho_{ij}^{(1)} + \Omega_s \rho_{ij}^{(1)} + ...$) to decouple Eq. (3), we derive the steady-state analytical solutions for $\rho_{31}$ and $\rho_{41}$ as follows:

$$\rho_{31} = \frac{2\Delta_p (i\Gamma + 2\Delta_p)\Omega_p + 4(-\Omega_{c1}\Omega_p + \Omega_{c2}\Omega_s)\Omega_{c1}^*}{(\Gamma - 2i\Delta_p)\left[(\Gamma - 2i\Delta_p)\Delta_p + 2i|\Omega_{c1}|^2 + 2i|\Omega_{c2}|^2\right]}, \quad (4)$$

$$\rho_{41} = \frac{2\Delta_p (i\Gamma + 2\Delta_p)\Omega_s + 4(\Omega_{c1}\Omega_p - \Omega_{c2}\Omega_s)\Omega_{c2}^*}{(\Gamma - 2i\Delta_p)\left[(\Gamma - 2i\Delta_p)\Delta_p + 2i|\Omega_{c1}|^2 + 2i|\Omega_{c2}|^2\right]}, \quad (5)$$

here, we assume that the detuning of the two strong control fields, $\Omega_{c1}$ and $\Omega_{c2}$, is zero, i.e., $\Delta_{c1} = \Delta_{c2} = 0$.

The propagation of the probe field $\Omega_p$ and the signal field $\Omega_s$ within the medium is governed by the Maxwell-Bloch equations, expressed as follows:

$$\left(\frac{1}{c}\frac{\partial}{\partial t} + \frac{\partial}{\partial z}\right)\Omega_p = i\frac{\Gamma_{31}\alpha_p}{2L}\rho_{31}, \quad (6)$$

$$\left(\frac{1}{c}\frac{\partial}{\partial t} + \frac{\partial}{\partial z}\right)\Omega_s = i\frac{\Gamma_{41}\alpha_s}{2L}\rho_{41}, \quad (7)$$

where $\alpha_j (j = p, s)$ represents the optical density of the medium. For simplicity, we assume $\alpha_p = \alpha_s = \alpha$. The parameter $L$ denotes the length of the medium. The phase-matching condition is satisfied in this case, expressed as $\Delta \mathbf{k} = \mathbf{k}_{c2} + \mathbf{k}_s - \mathbf{k}_{c1} - \mathbf{k}_p = 0$.

By substituting Eqs. (4) and (5) into Eqs. (6) and (7), and applying the boundary conditions $\Omega_p(z=0) = \Omega_p(0)$ and $\Omega_s(z=0) = 0$. The steady-state solutions for the probe field $\Omega_p$ and the signal field $\Omega_s$ propagating in the medium are given by:

$$\Omega_p(z) = \frac{\Omega_p(0)}{|\Omega_{c1}|^2 + |\Omega_{c2}|^2}\left[|\Omega_{c1}|^2 \exp\left(-\frac{\alpha\Gamma}{2\Gamma - 4i\Delta_p} \cdot \frac{z}{L}\right) + |\Omega_{c2}|^2 \exp\left(-\frac{\alpha\Gamma\Delta_p}{2\Gamma\Delta_p - 4i\Delta_p^2 + 4i|\Omega_{c1}|^2 + 4i|\Omega_{c2}|^2} \cdot \frac{z}{L}\right)\right], \quad (8)$$

$$\Omega_s(z) = \frac{\Omega_p(0)\Omega_{c1}\Omega_{c2}^*}{|\Omega_{c1}|^2 + |\Omega_{c2}|^2}[-\exp\left(-\frac{\alpha\Gamma}{2\Gamma - 4i\Delta_p}\cdot\frac{z}{L}\right) + \\ \exp\left(-\frac{\alpha\Gamma\Delta_p}{2\Gamma\Delta_p - 4i\Delta_p^2 + 4i|\Omega_{c1}|^2 + 4i|\Omega_{c2}|^2}\cdot\frac{z}{L}\right)].$$  (9)

Before analyzing the theoretical results, we first introduce the most common vortex beams, which take the form of Laguerre-Gaussian (LG) modes. The complex amplitude distribution of a LG beam can be expressed in a cylindrical coordinate system [54,67]:

$$\Omega(r,\varphi) = \Omega_0 \frac{1}{\sqrt{|l|!}}\left(\frac{\sqrt{2}r}{w_0}\right)^{|l|} L_p^{|l|}\left(\frac{2r^2}{w_0^2}\right) e^{-\frac{r^2}{w_0^2}} e^{il\varphi},$$  (10)

where $\Omega_0$, $r$, $w_0$, $l$, and $p$ denote the intensity of the LG beam, the radial coordinate, the beam waist, the topological charge, and the radial index, respectively. $L_p^{|l|}$ represents the associated Laguerre polynomial, expressed as:

$$L_p^{|l|}(x) = \frac{e^x x^{-|l|}}{p!}\frac{d^p}{dx^p}\left(x^{|l|+p} e^{-x}\right),$$  (11)

where $x = 2r^2/w_0^2$ governs the radial dependence of the LG beam for different radial indices. When $l \neq 0$, the LG beam carries OAM aligned with the optical axis.

**B.1. Only One Input Beam Carries OAM**

For the sake of clarity, we assume that only the probe field $\Omega_p(0)$ is an LG beam at the medium entrance ($z=0$), which is given by Eq. (10). As indicated by Eqs. (8) and (9), the generated signal field $\Omega_s(z)$ inherits the same OAM as the incident probe field $\Omega_p(0)$. We next examine the influence of two strong control fields, $\Omega_{c1}$ and $\Omega_{c2}$, on the vortex OAM transfer between the probe and signal fields. Figs. 2(a)-(c) illustrate the evolution of the normalized intensities, $|\Omega_p(z)|^2/|\Omega_p(0)|^2$ and $|\Omega_s(z)|^2/|\Omega_p(0)|^2$, as functions of the dimensionless length $z/L$ under varying control field strengths. As the interaction depth between the light field and the medium increases, a portion of the energy from the initial probe field is progressively transferred to the generated signal field via the FWM process. Simultaneously, this process entails the transfer of OAM, which ultimately stabilizes. Specifically, Fig. 2(a) presents the energy transfer from the probe field to the generated signal field for control field parameters $\Omega_{c1} = 1\Gamma$ and $\Omega_{c2} = 2\Gamma$. It is evident that, under these conditions, the probe field experiences minimal loss, while the generated signal field remains weak. Fig. 2(b) illustrates the energy exchange between the probe and signal fields for $\Omega_{c1} = 1\Gamma$ and $\Omega_{c2} = 1\Gamma$. At this condition, the energy conversion efficiency is maximized, and both fields attain equal intensity at equilibrium. Fig. 2(c) depicts the energy transfer dynamics for $\Omega_{c1} = 2\Gamma$ and $\Omega_{c2} = 1\Gamma$. Under these conditions, compared to Fig. 2(a), the probe field undergoes substantial loss during the transfer process, nearly depleting its energy. Despite its low conversion efficiency, the generated signal field exhibits the same characteristics as in Fig. 2(a). Evidently, maintaining equal Rabi frequencies for the two control fields ($\Omega_{c1} = \Omega_{c2}$) optimizes vortex optical transmission efficiency. This phenomenon arises from the competition between absorption and gain. From Eqs. (4)–(7), it follows that when $\Omega_{c1}\Omega_p = \Omega_{c2}\Omega_s$, the equations fully decouple, leading to independent evolution and an eventual equilibrium state. Specifically, for $\Omega_{c1} < \Omega_{c2}$, equilibrium requires $\Omega_p > \Omega_s$; for $\Omega_{c1} > \Omega_{c2}$, equilibrium necessitates $\Omega_p < \Omega_s$; and for $\Omega_{c1} = \Omega_{c2}$, equilibrium is achieved with $\Omega_p = \Omega_s$. In this latter case, the energy conversion efficiency reaches its maximum, as the competing processes are perfectly balanced.

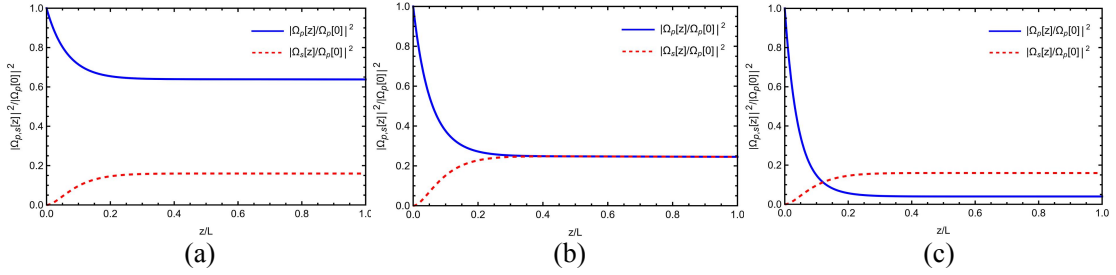

(a)　　　　　　　　　　(b)　　　　　　　　　　(c)

**Fig. 2.** Evolution of the probe field intensity $|\Omega_p(z)|^2/|\Omega_p(0)|^2$ and the generated field intensity $|\Omega_s(z)|^2/|\Omega_p(0)|^2$ as a function of the dimensionless propagation length $z/L$ for different control field intensities. (a) $\Omega_{c1} = 1\Gamma$, $\Omega_{c2} = 2\Gamma$; (b) $\Omega_{c1} = 1\Gamma$, $\Omega_{c2} = 1\Gamma$; (c) $\Omega_{c1} = 2\Gamma$, $\Omega_{c2} = 1\Gamma$. The remaining parameters in all three cases are $\Delta_p = 0.1$ and $\alpha = 30$.

Next, we examine the effect of detuning on conversion efficiency under optimal energy transfer conditions, i.e., when $\Omega_{c1} = \Omega_{c2}$. Fig. 3 illustrates the evolution of the normalized intensities of the probe and signal fields as a function of the dimensionless propagation length $z/L$ for $\Delta_p = 0, 0.5\Gamma, 1\Gamma$. As the detuning $\Delta_p$ increases, energy transfer from the probe field to the signal field is progressively attenuated, due to the fact that at exact resonance ($\Delta_p = 0$), the system evolves into a dark state as the optical field propagates through the medium, thereby establishing EIT and minimizing loss. When $\Delta_p$ deviates from resonance, this dark state is disrupted, leading to significant dissipation.

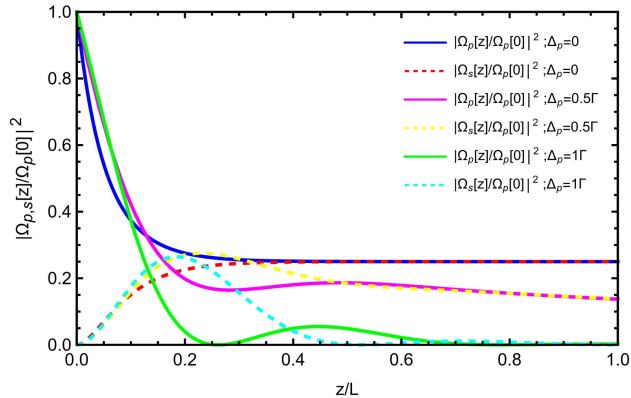

**Fig. 3.** Evolution of the normalized intensities of the probe and signal fields as a function of the dimensionless propagation length $z/L$ for detuning values $\Delta_p = 0, 0.5\Gamma, 1\Gamma$. The remaining parameters are $\Omega_{c1} = 1\Gamma$, $\Omega_{c2} = 1\Gamma$, and $\alpha = 30$.

To further elucidate the vortex light transfer process, we present the intensity and phase distributions of the generated signal field for different LG modes of the probe field $\Omega_p(0)$. First, we consider the case of exact resonance ($\Delta_p = 0$) in the system. Fig. 4 illustrates the intensity and phase distributions of the generated signal field $\Omega_s$ at $z = L$, for different $LG_p^l$ modes of the probe field $\Omega_p(0)$, while the two control fields are conventional light, with $\Omega_{c1} = \Omega_{c2} = 1\Gamma$. Specifically, for $p_p = 0$ and $l_p = -2, -1, 1, 2$, the signal field exhibits a single-ring intensity structure with a phase singularity at the dark center, where the phase changes by $2\pi l_p$ around the singularity. $\pm l_p$ indicates that the chiral direction of the spiral phase is opposite, and the larger the value of $|l_p|$, the bigger the dark hole at the center. For $p_p \neq 0$ and $l_p = -2, -1, 1, 2$, the signal field intensity exhibits a structure of $p_p + 1$ rings, with the corresponding phase distribution containing $p_p + 1$ phase regions, each phase region undergoes $|l_p|$ complete phase change cycles. Numerical results confirm that the generated signal field $\Omega_s$ retains the same vortex structure as the input probe field $\Omega_p(0)$, demonstrating that the OAM information of the vortex probe field is faithfully transferred through the FWM.

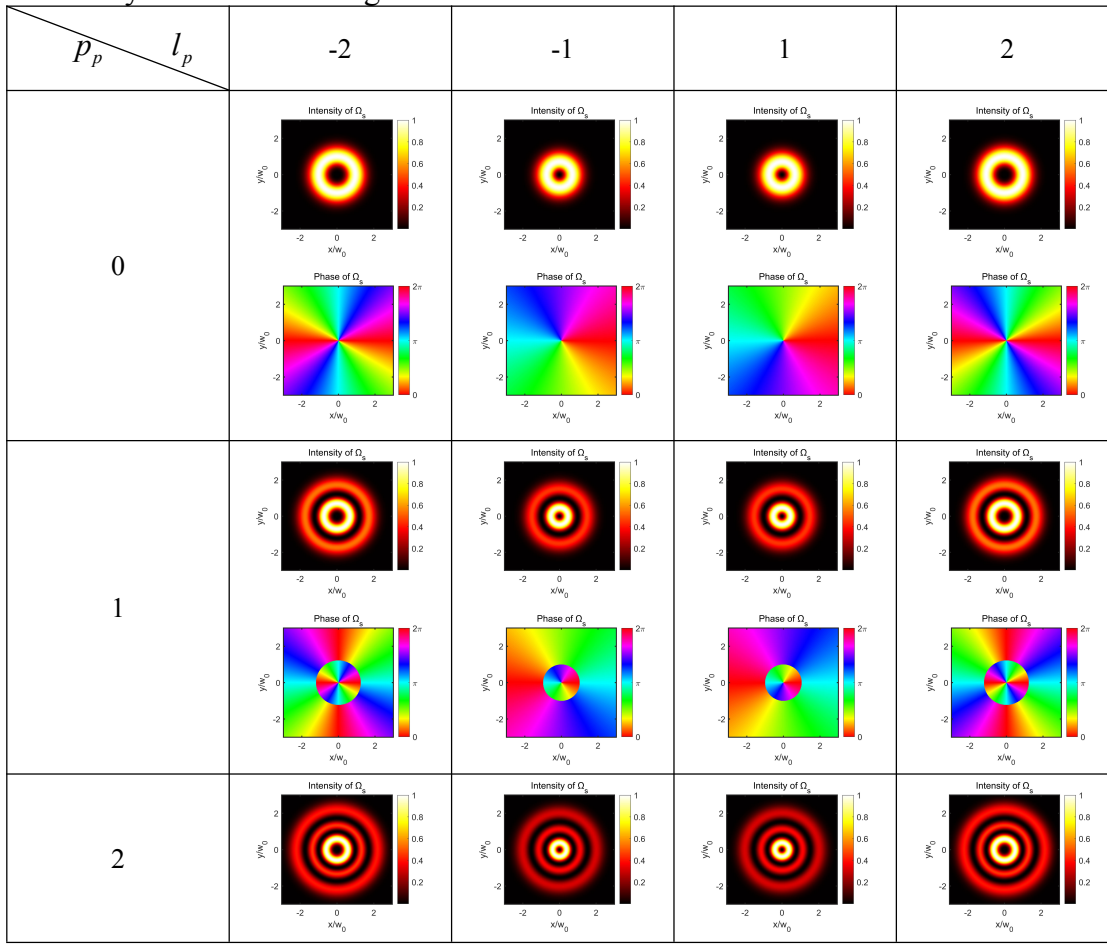

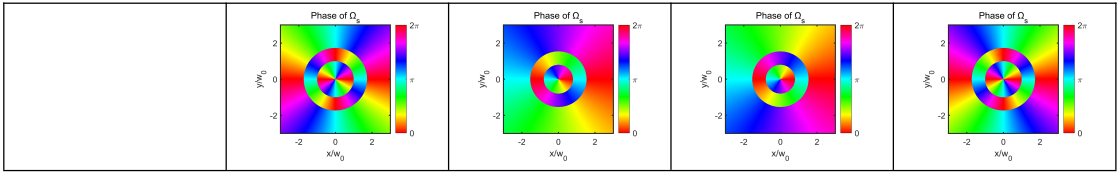

**Fig. 4.** Intensity and phase distributions of the generated signal field $\Omega_s$ at exact resonance ($\Delta_p = 0$) for different $LG_p^l$ modes of the probe field $\Omega_p(0)$. The other parameters are $\alpha = 30$, $\Omega_{p0} = 0.01\Gamma$, $\Omega_{c1} = 1\Gamma$, and $\Omega_{c2} = 1\Gamma$.

Next, we examine the scenario in which the system exhibits detuning ($\Delta_p = 0.001\Gamma$). Fig. 5 illustrates the intensity and phase distributions of the generated signal field $\Omega_s$ under varying $LG_p^l$ modes of the probe field $\Omega_p(0)$. Compared to Fig. 4, the presence of detuning induces varying degrees of phase distortion, along with significant attenuation of the outer-ring intensity distribution. This phenomenon arises because detuning induces strong refraction and dissipation in the medium.

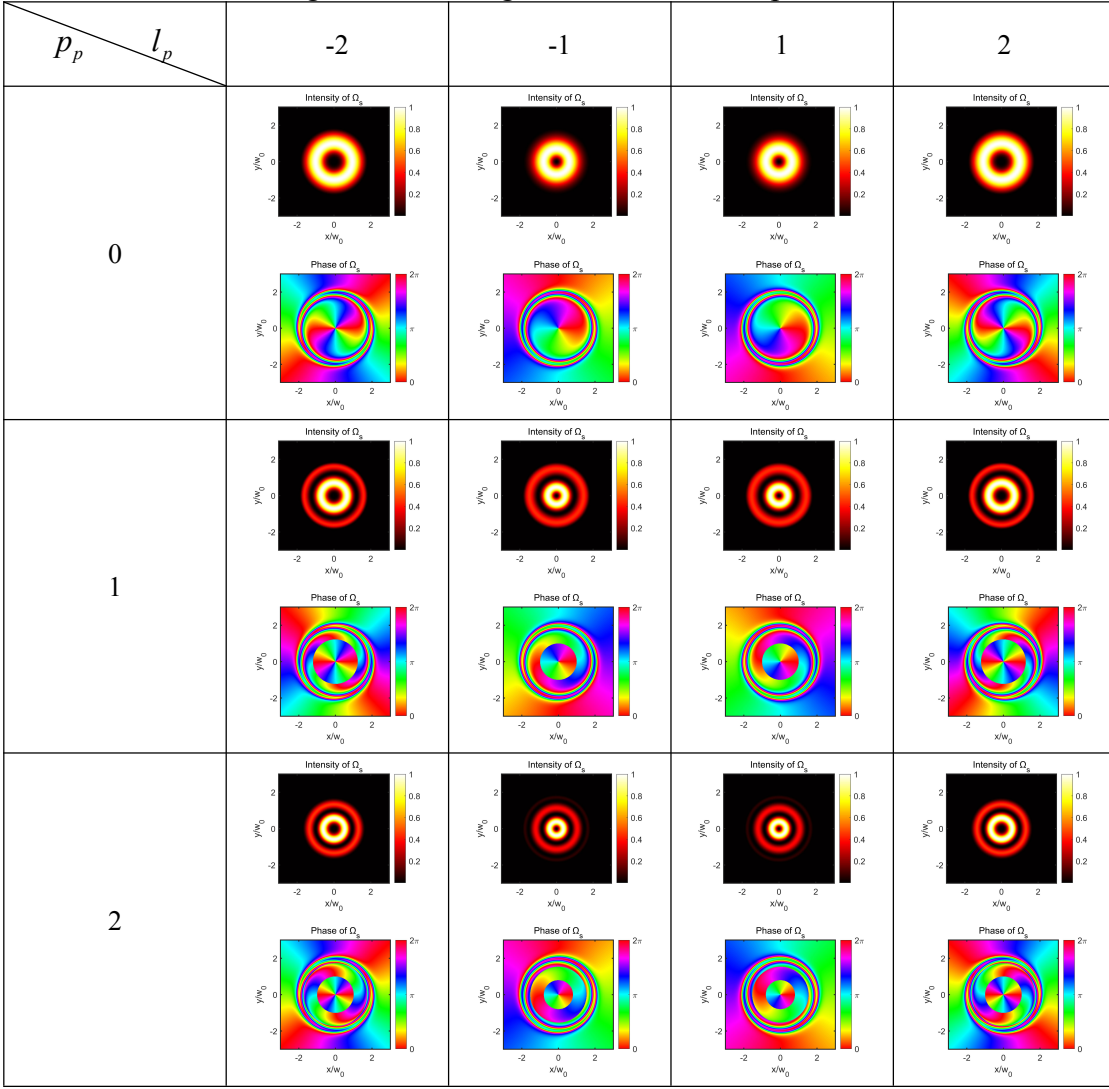

**Fig. 5.** Intensity and phase distributions of the generated signal field $\Omega_s$ for a detuning of $\Delta_p = 0.001\Gamma$, with the probe field $\Omega_p(0)$ in different $LG_p^l$ modes. The other parameters are consistent with those in Fig. 4.

## B.2. Two Beams of Light in the Input Field Carry OAM

For clarity in our discussion, we assume that the probe field $\Omega_p$ and the control field $\Omega_{c1}$ are initially vortex beams, whereas the control field $\Omega_{c2}$ remains a conventional light field. In this analysis, we focus solely on the case where the radial index $p = 0$. Based on Eqs. (8) and (9), the generated signal field $\Omega_s(z)$ acquires the combined OAM of the incident probe field $\Omega_p(0)$ and the control field $\Omega_{c1}$, such that $l_s = l_p + l_{c1}$. Fig. 6 illustrates the intensity and phase distributions of the generated signal field for different $LG_0^l$ mode combinations of the probe field $\Omega_p$ and the control field $\Omega_{c1}$. Specifically, we examine the cases where the probe field carries $l_p = -2, -1, 1, 2$ and the control field carries $l_{c1} = -2, -1, 1, 2$. The results indicate that the intensity and phase of the generated signal field are modulated by the two vortex fields, $\Omega_p$, $\Omega_{c1}$. The intensity of the signal field exhibits a donut-shaped profile, with the central dark region expanding as $|l|$ increases. The OAM of the signal field follows $l_s = l_p + l_{c1}$, as evidenced by the phase distribution, where the phase shift is $2\pi l_s$. Notably, in the negative diagonal region of Fig. 6, while the signal field maintains an annular intensity profile, it lacks a helical phase structure and, therefore, no longer qualifies as vortex light. This phenomenon arises because, despite the total topological charge summing to $l_s = 0$, the $LG$ mode retains intensity modulation dictated by the $\left(\sqrt{2}r/w_0\right)^{|l|}$ term.

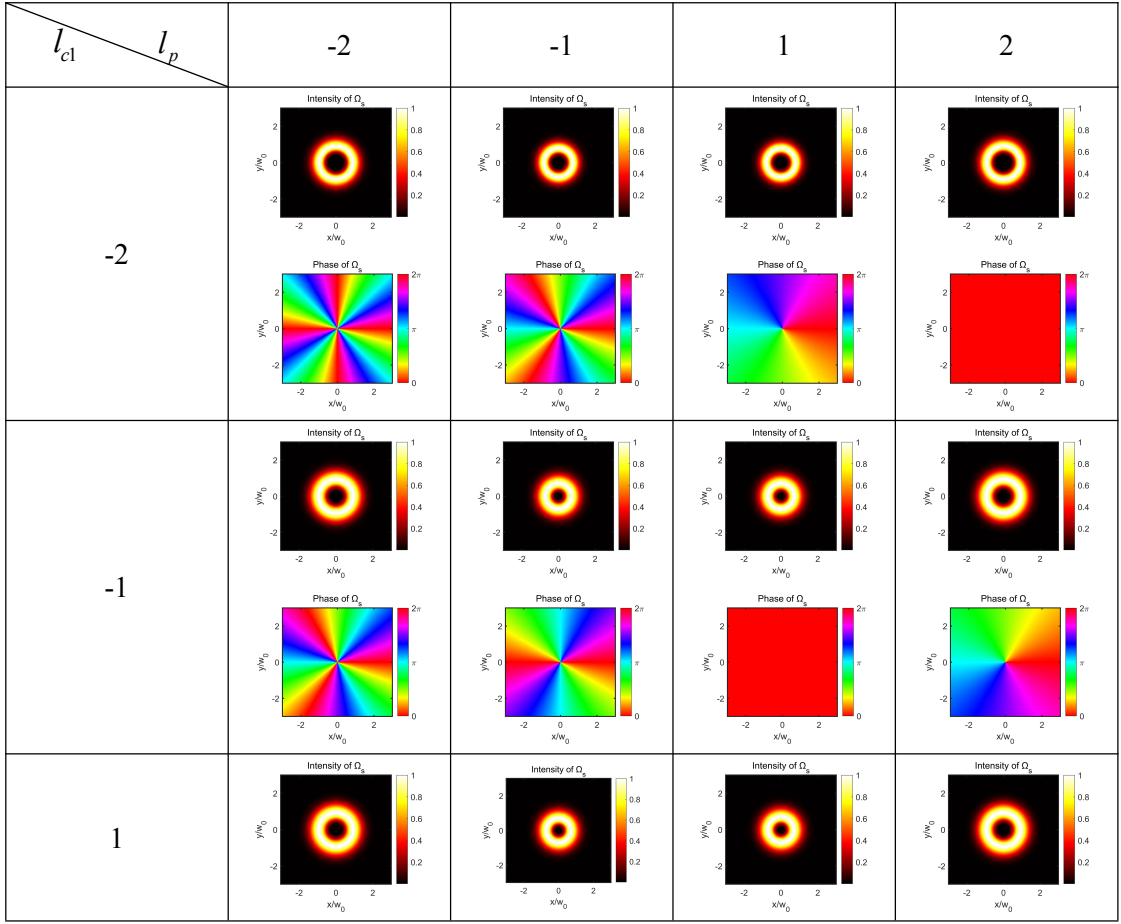

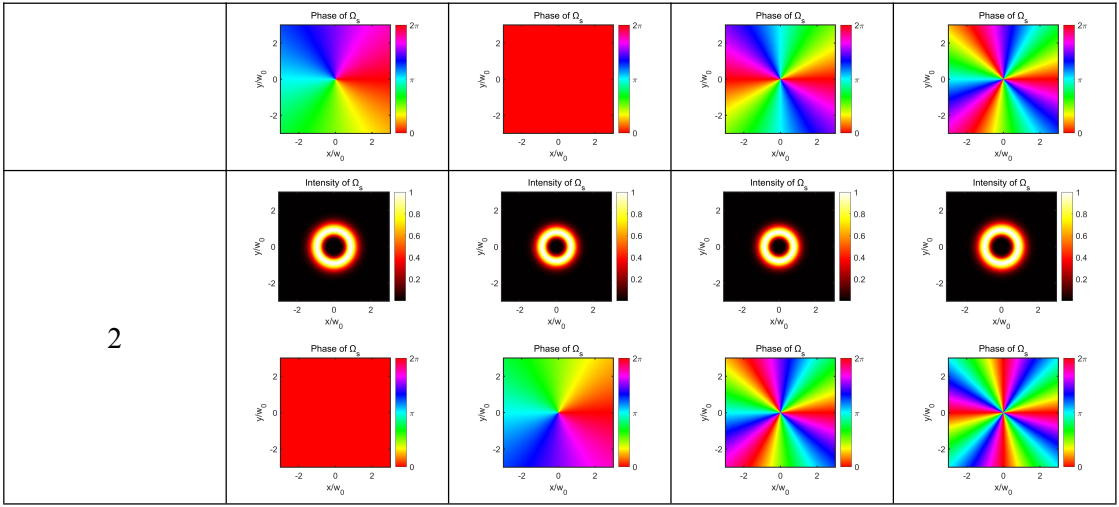

**Fig. 6.** Intensity and phase distributions of the generated signal field for different $LG_0^l$ mode combinations of the probe field $\Omega_p$ and the control field $\Omega_{c1}$ (i.e., $l_p = -2, -1, 1, 2$, $l_{c1} = -2, -1, 1, 2$). The other parameters are $\Delta_p = 0$, $\alpha = 30$, $\Omega_{p0} = 0.01\Gamma$, $\Omega_{c10} = 1\Gamma$, and $\Omega_{c2} = 1\Gamma$.

### B.3. All Three Beams of Light in the Input Field Carry OAM

In the following, we examine the more prevalent scenario, where all three beams of light in the incident field possess OAM, meaning that $\Omega_{c2}$ is also considered to exhibit vortex light properties. For the purpose of this discussion, we focus solely on the case where the radial index is set to $p = 0$. As described by Eqs (8) and (9), the OAM responsible for generating the signal field $\Omega_s(z)$ is modulated by the incident probe field $\Omega_p(0)$ and the control fields $\Omega_{c1}$ and $\Omega_{c2}$, with their respective topological charges satisfying the relation $l_s = l_p + l_{c1} - l_{c2}$. Previous studies have primarily considered scenarios where either the probe field $\Omega_p$ or the control field $\Omega_{c1}$ carries OAM, without accounting for the possibility that the control field $\Omega_{c2}$ itself may also exhibit vortex light properties. This omission may stem from the fact that the control field $\Omega_{c2}$, appearing on the right-hand side of Eq. (9), is complex conjugated, and is thus assumed to lack any physical significance. However, experimental reports [50] have demonstrated that all three input light beams can carry OAM, thus making it plausible to consider the possibility that the control field $\Omega_{c2}$ may also carry OAM. In this study, we conduct a theoretical analysis of this scenario. We examine the following three scenarios: a) The topological charge of the control field $\Omega_{c2}$ is fixed at $l_{c2} = 1$, while the probe field $\Omega_p$ and control field $\Omega_{c1}$ each adopt distinct $LG_0^l$ modes, with $l_p = -2, -1, 1, 2$ and $l_{c1} = -2, -1, 1, 2$. b) The topological charge of the control field $\Omega_{c1}$ is fixed at $l_{c1} = 1$, while the probe field $\Omega_p$ and control field $\Omega_{c2}$ each adopt distinct $LG_0^l$ modes, with $l_p = -2, -1, 1, 2$ and $l_{c2} = -2, -1, 1, 2$. c) The topological charge of the probe field $\Omega_p$ is fixed at $l_p = 1$, while the control fields $\Omega_{c1}$ and $\Omega_{c2}$ each adopt distinct $LG_0^l$ modes, with $l_{c1} = -2, -1, 1, 2$ and $l_{c2} = -2, -1, 1, 2$. The intensity and phase distributions of the generated signal field $\Omega_s$ for each of these three cases are presented in Fig. 7(a), 7(b), and 7(c), respectively. Specifically, the intensity distribution of the generated signal field is modulated by the three incident fields, forming a donut-shaped structure. Additionally, the dark center increases as $|l|$ increases. The phase distribution of the signal field clearly reveals that it is modulated by the OAM of the three incident fields, with the topological charge satisfying the relation $l_s = l_p + l_{c1} - l_{c2}$. This suggests that during the FWM process, the OAM of the incident and signal fields can undergo both additive and subtractive operations. These findings may offer meaningful insights for potential applications in quantum information encoding, quantum computing, and related areas.

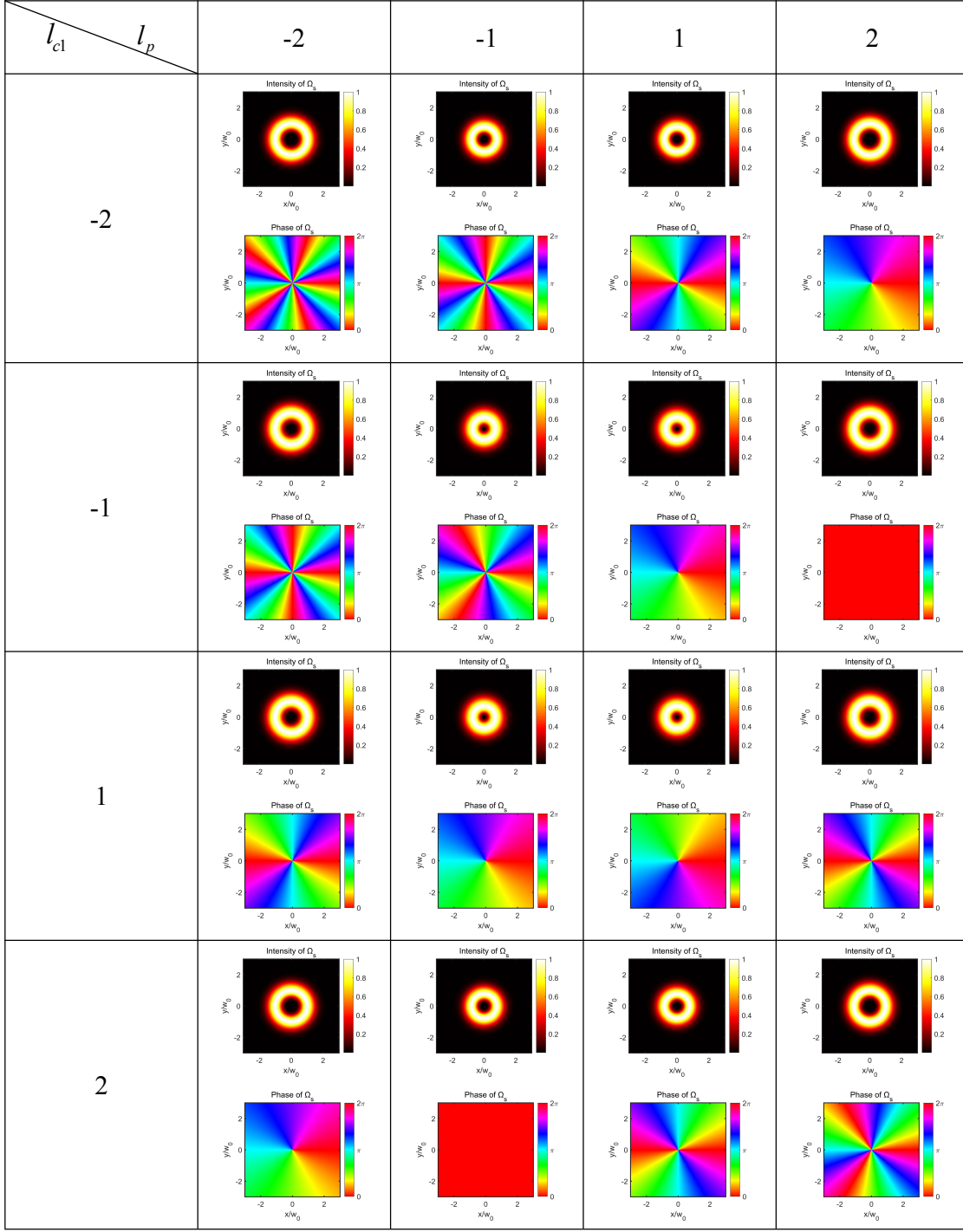

**Fig. 7(a).** When the topological charge of the control field $\Omega_{c2}$ is fixed at $l_{c2}=1$, and the probe field $\Omega_p$ and control field $\Omega_{c1}$ adopt distinct $LG_0^l$ modes (i.e., $l_p=-2,-1,1,2$, $l_{c1}=-2,-1,1,2$), the intensity and phase distributions of the signal field are generated. The other parameters are $\Delta_p=0$, $\alpha=30$, $\Omega_{p0}=0.01\Gamma$, $\Omega_{c10}=1\Gamma$, and $\Omega_{c20}=1\Gamma$.

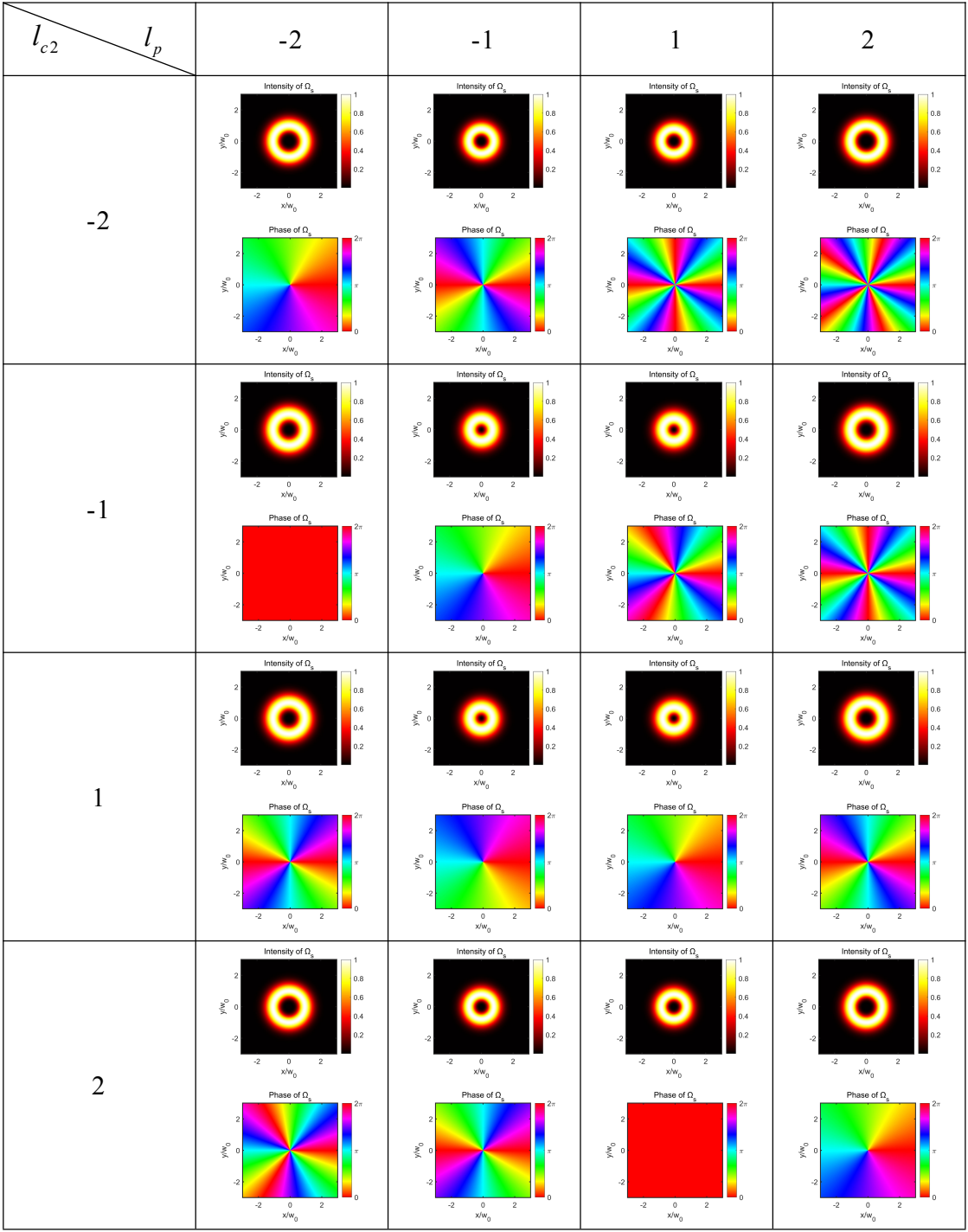

**Fig. 7(b).** When the topological charge of the control field $\Omega_{c1}$ is fixed at $l_{c1}=1$, and the probe field $\Omega_p$ and control field $\Omega_{c2}$ adopt distinct $LG_0^l$ modes (i.e., $l_p=-2,-1,1,2$, $l_{c2}=-2,-1,1,2$), the intensity and phase distributions of the signal field are generated. The other parameters are consistent with those in 7(a).

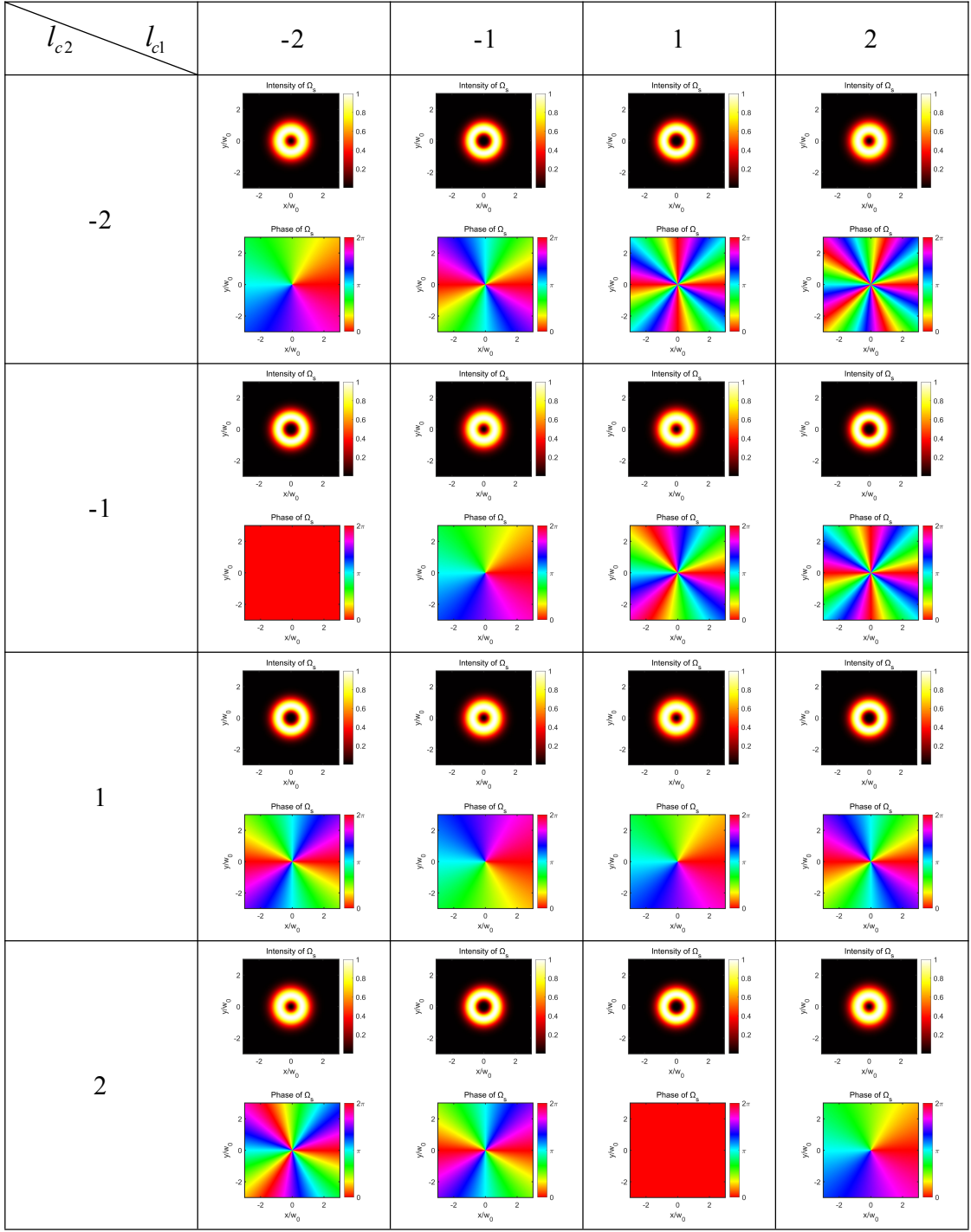

**Fig. 7(c).** When the topological charge of the probe field $\Omega_p$ is fixed at $l_p = 1$, and the control fields $\Omega_{c1}$ and $\Omega_{c2}$ adopt distinct $LG_0^l$ modes (i.e., $l_{c1} = -2, -1, 1, 2$, $l_{c2} = -2, -1, 1, 2$), the intensity and phase distributions of the signal field are generated. The other parameters are consistent with those in 7(a).

## C. Manipulation of Slow and Fast Light

The control of group velocities of the probe and signal fields through two strong control fields in the FWM process—enabling the conversion between slow and fast light within the medium—remains unexplored in dual-Λ-type systems. To more comprehensively analyze the group velocities of both the probe and signal fields during the FWM process, we also consider the relative phase between these fields. By introducing phase, the Rabi frequencies of the probe and signal fields can be expressed as $\Omega_p = |\Omega_p| e^{i\phi_p}$ and $\Omega_s = |\Omega_s| e^{i\phi_s}$, where $\phi_p$ and $\phi_s$ represent the respective phases of the corresponding light fields. Therefore, the polarization rate $\chi$ of the input probe field and the generated signal field can be expressed as:

$$\chi_p(\Delta) = \frac{N|\mu_{31}|^2}{\varepsilon_0 \hbar} \frac{\rho_{31}}{\Omega_p} = \frac{N|\mu_{31}|^2}{\varepsilon_0 \hbar} \frac{2\Delta_p (i\Gamma + 2\Delta_p) + 4\left(-\Omega_{c1} + \Omega_{c2} \frac{\Omega_s}{\Omega_p} e^{-i\phi}\right)\Omega_{c1}^*}{(\Gamma - 2i\Delta_p)\left[(\Gamma - 2i\Delta_p)\Delta_p + 2i|\Omega_{c1}|^2 + 2i|\Omega_{c2}|^2\right]}, \quad (12)$$

$$\chi_s(\Delta) = \frac{N|\mu_{41}|^2}{\varepsilon_0 \hbar} \frac{\rho_{41}}{\Omega_s} = \frac{N|\mu_{41}|^2}{\varepsilon_0 \hbar} \frac{2\Delta_p (i\Gamma + 2\Delta_p) + 4\left(-\Omega_{c2} + \Omega_{c1} \frac{\Omega_p}{\Omega_s} e^{i\phi}\right)\Omega_{c2}^*}{(\Gamma - 2i\Delta_p)\left[(\Gamma - 2i\Delta_p)\Delta_p + 2i|\Omega_{c1}|^2 + 2i|\Omega_{c2}|^2\right]}, \quad (13)$$

where we assume that the probe field is equivalent to the electric dipole matrix elements of the signal field, i.e., $\mu_{31} = \mu_{41} = \mu$; $\phi = \phi_p - \phi_s$ represents the relative phase between the probe and signal fields; $N$ denotes the atomic number density, and $\varepsilon_0$ is the dielectric constant in a vacuum.

The real and imaginary components of the polarization $\chi$ correspond to the dispersion and absorption characteristics of the medium, respectively. By examining how the system's tunable parameters affect the medium's dispersion properties, we can investigate both superluminal (fast light) and subluminal (slow light) propagation of the probe and signal fields. Normal dispersion leads to subluminal propagation, whereas anomalous dispersion gives rise to superluminal propagation. Below, we will investigate how the Rabi frequency of the two strong control fields and relative phase $\phi$ between the probe and signal fields influence the dispersion properties of the medium, thereby affecting the group velocities of both the probe and signal fields.

**C.1. Control of the Group Velocity of the Probe Field**

Fig. 8 illustrates the response characteristics of the medium to the absorption and dispersion of the probe field when the two control fields assume different values and exhibit varying relative phases $\phi$ between the probe and signal fields. Figs. 8(a)-(e) present the absorption and dispersion curves of the probe field for the following control field configurations: $|\Omega_{c1}| = 1\Gamma$, $|\Omega_{c2}| = 5\Gamma$; $|\Omega_{c1}| = 1\Gamma$, $|\Omega_{c2}| = 2\Gamma$; $|\Omega_{c1}| = 1\Gamma$, $|\Omega_{c2}| = 1\Gamma$; $|\Omega_{c1}| = 2\Gamma$, $|\Omega_{c2}| = 1\Gamma$; and $|\Omega_{c1}| = 5\Gamma$, $|\Omega_{c2}| = 1\Gamma$. In this context, for each pair of control fields, we also examine the impact of $\phi = 2\pi, 3\pi/2, \pi, \pi/2$. Specifically, when $|\Omega_{c1}| = |\Omega_{c2}| = 1\Gamma$, three absorption peaks are observed in the probe field $\Omega_p$, and the slope of the dispersion curve in the absorption peak region is negative, indicating superluminal propagation. Conversely, the dispersion curve slope in the region between the absorption peaks is positive, indicating subluminal propagation, as depicted in Fig. 8(c). When $|\Omega_{c1}| = 1\Gamma$ is fixed and $|\Omega_{c2}|$ is gradually increased, the intermediate absorption peak of the probe field gradually decreases, while the two side absorption peaks increase. This implies that the slope of the dispersion curve in the intermediate absorption peak region transitions from negative to positive, and the corresponding superluminal propagation gradually shifts from subluminal propagation. When $|\Omega_{c2}| = 1\Gamma$ is fixed and $|\Omega_{c1}|$ is gradually increased, the situation reverses as described above. The intermediate absorption peak of the probe field increases, while the two side absorption peaks decrease, indicating that the anomalous dispersion in the absorption peak regions on both sides gradually transitions to normal dispersion. Consequently, the superluminal propagation transitions to subluminal propagation. The relative phase $\phi$ has minimal impact on the absorption and dispersion of the probe field.

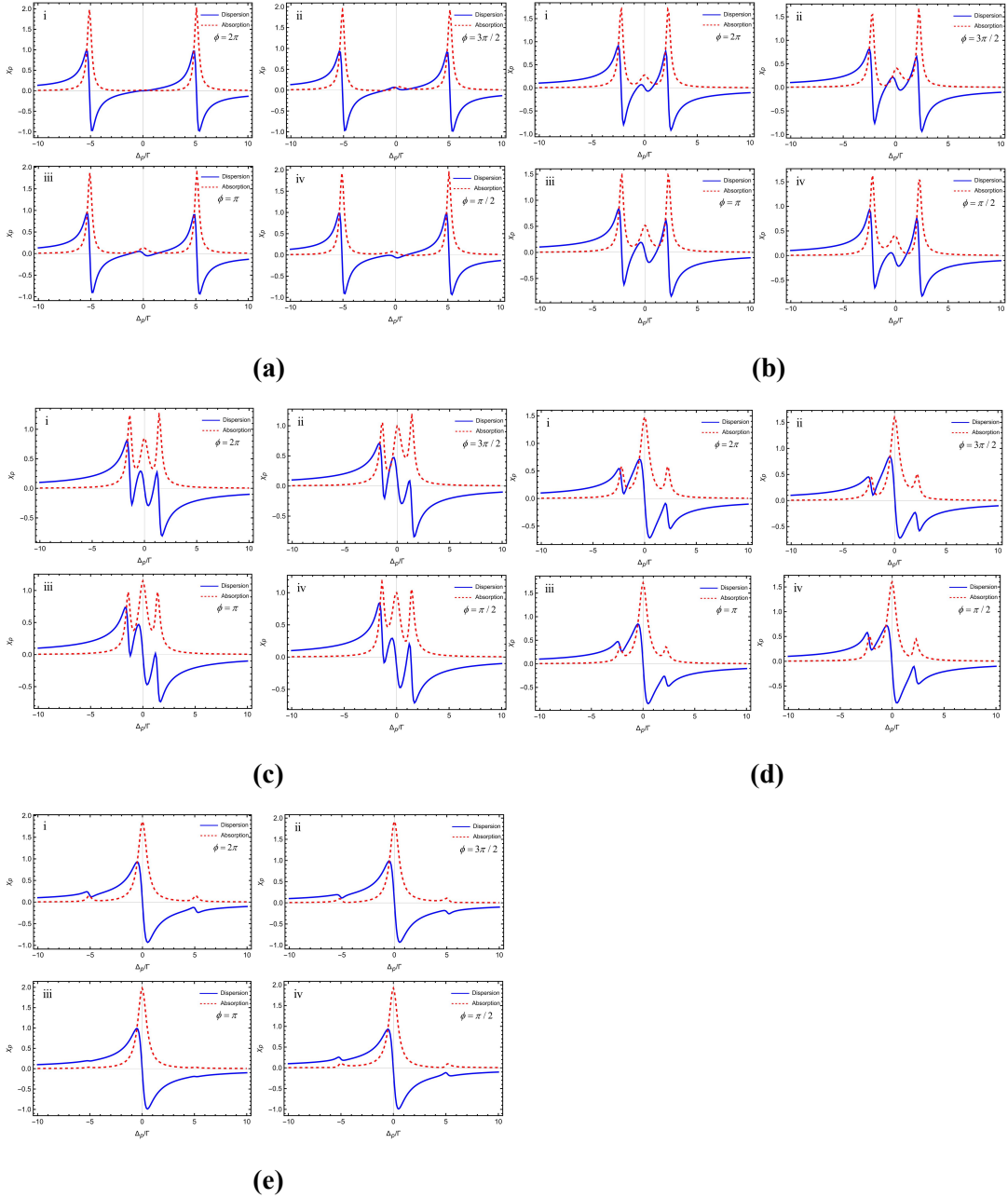

**Fig. 8.** Dependence of the real (solid line) and imaginary (dashed line) parts of the probe field polarization rate $\chi_p$ on $\Delta_p/\Gamma$, expressed in units of $N|\mu_{31}|^2/\varepsilon_0\hbar$, under varying strong control field values. Each subplot illustrates the behavior for different relative phase values ($\phi = 2\pi, 3\pi/2, \pi, \pi/2$). (a) $|\Omega_{c1}| = 1\Gamma$, $|\Omega_{c2}| = 5\Gamma$. (b) $|\Omega_{c1}| = 1\Gamma$, $|\Omega_{c2}| = 2\Gamma$. (c) $|\Omega_{c1}| = 1\Gamma$, $|\Omega_{c2}| = 1\Gamma$. (d) $|\Omega_{c1}| = 2\Gamma$, $|\Omega_{c2}| = 1\Gamma$. (e) $|\Omega_{c1}| = 5\Gamma$, $|\Omega_{c2}| = 1\Gamma$. Other parameters are set as $|\Omega_p| = 0.2\Gamma$ and $|\Omega_s| = 0.03\Gamma$.

## C.2. Control of the Group Velocity of the Generated Signal Field

Fig. 9 presents the response characteristics of the medium in terms of signal field absorption and dispersion when the Rabi frequencies of the two control fields and the relative phase $\phi$ are varied. Figs. 9(a)-(e) show the absorption and dispersion curves of the signal field for the following control field parameters: $|\Omega_{c1}|=1\Gamma$, $|\Omega_{c2}|=5\Gamma$; $|\Omega_{c1}|=1\Gamma$, $|\Omega_{c2}|=2\Gamma$; $|\Omega_{c1}|=1\Gamma$, $|\Omega_{c2}|=1\Gamma$; $|\Omega_{c1}|=2\Gamma$, $|\Omega_{c2}|=1\Gamma$; and $|\Omega_{c1}|=5\Gamma$, $|\Omega_{c2}|=1\Gamma$, respectively. For each set of parameter values, we also consider the influence of the relative phase $\phi = 2\pi, 3\pi/2, \pi, \pi/2$. Specifically, when $|\Omega_{c1}|=|\Omega_{c2}|=1\Gamma$, the subplot i of Fig. 8(c) reveals a gain peak and two absorption peaks on either side near $\Delta_s = 0$, which correspond to subluminal and superluminal propagation, respectively. In subplot ii of Fig. 8(c), an absorption peak and a gain peak are observed near the $\Delta_s = 0$ of the signal field, corresponding to superluminal and subluminal propagation, respectively, with a sharp transition in the group velocity. The behavior observed in subplots iii and iv of Fig. 8(c) is the inverse of that in subplots i and ii, respectively. When $|\Omega_{c1}|=1\Gamma$ is held constant while $|\Omega_{c2}|$ is gradually increased, or vice versa, the absorption and dispersion profiles of the signal field progressively broaden. It is noteworthy that, unlike in the probe field, the relative phase plays a significant role in modulating the results of the signal field.

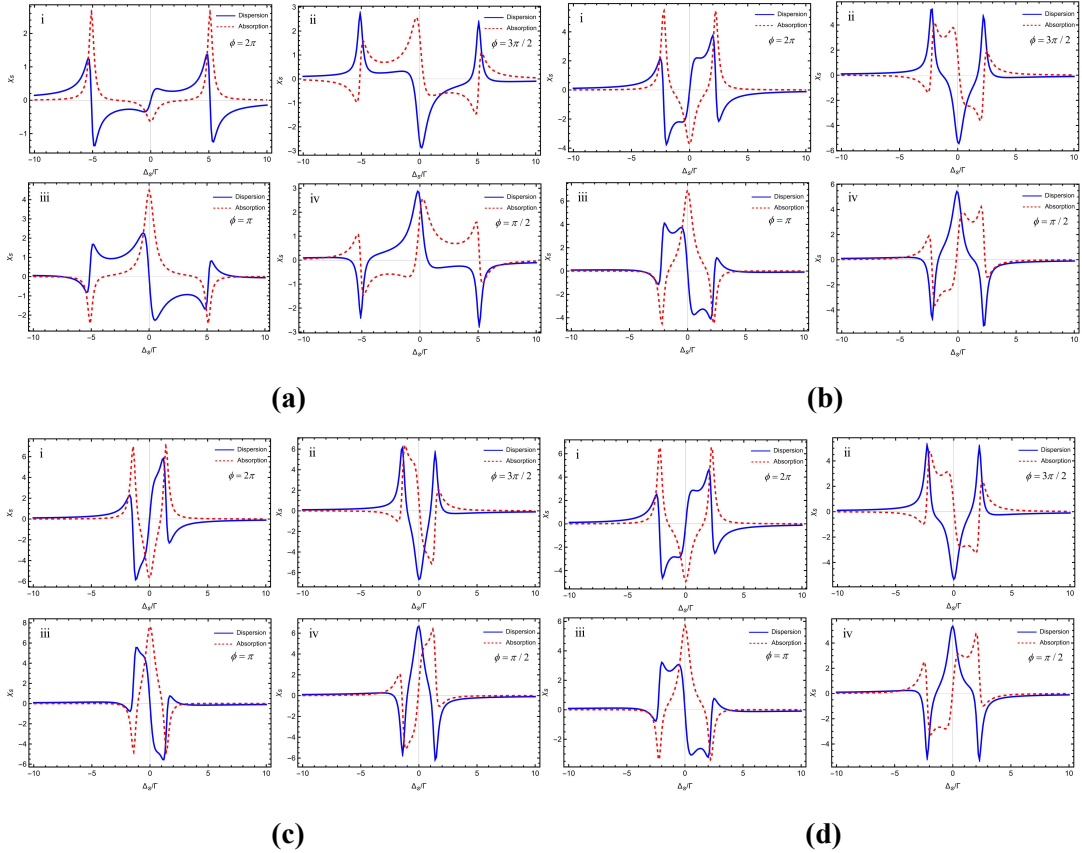

(a)　　　　　　　　(b)

(c)　　　　　　　　(d)

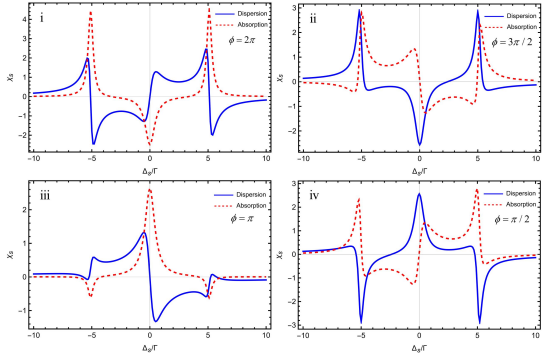

**(e)**

**Fig. 9.** Dependence of the real (solid line) and imaginary (dashed line) parts of the signal field polarization rate $\chi_s$ on $\Delta_s/\Gamma$, expressed in units of $N|\mu_{41}|^2/\varepsilon_0\hbar$, under varying strong control field values. Each subplot illustrates the behavior for different relative phase values ($\phi = 2\pi, 3\pi/2, \pi, \pi/2$). (a) $|\Omega_{c1}| = 1\Gamma$, $|\Omega_{c2}| = 5\Gamma$. (b) $|\Omega_{c1}| = 1\Gamma$, $|\Omega_{c2}| = 2\Gamma$. (c) $|\Omega_{c1}| = 1\Gamma$, $|\Omega_{c2}| = 1\Gamma$. (d) $|\Omega_{c1}| = 2\Gamma$, $|\Omega_{c2}| = 1\Gamma$. (e) $|\Omega_{c1}| = 5\Gamma$, $|\Omega_{c2}| = 1\Gamma$. The other parameters are consistent with those in Fig. 8.

Thus, by manipulating the Rabi frequency intensities of the two control fields and the relative phase angle $\phi$ between the probe and signal fields, we can modulate the group velocity of both the probe and signal fields, enabling the switching between superluminal and subluminal propagation.

## 3. Conclusion

In this paper, we investigate the transfer of OAM from the input vortex field in the four-level dual-Λ system to the generated signal field via the FWM process, where the OAMs of the involved fields obey a specific algebraic relationship. Furthermore, we examine the group velocities of both the probe and signal fields during the FWM process, revealing that switching between matched vortex slow and fast light can be realized by tuning the two strong control fields and the relative phase $\phi$ between the probe and signal fields. This research provide a significant extension and supplement of prior studies, as outlined below:

We begin by analyzing the transfer of OAM in vortex light, specifically distinguishing between cases where one, two, or all three input light fields carry OAM. The first two scenarios have been partially explored in previous studies. Our contribution differs in two key aspects: (1) We elucidate the physical conditions necessary for achieving efficient vortex light transmission. Prior studies commonly assumed equal control fields ($\Omega_{c1} = \Omega_{c2}$), without addressing the underlying physical mechanisms through which variations in the control fields influence transmission efficiency. We offer a detailed theoretical explanation of these mechanisms in Section B.1 of the paper. (2) We analyze the influence of detuning on both the transmission efficiency and phase distortion in the OAM transfer process, highlighting its role in modifying the refractive index and dissipative properties of the medium. In the third case, we focus on the configuration in which all three input fields are vortex beams propagating through the medium, and clarify that previous studies may have overlooked the possibility of the control field $\Omega_2$ also carrying OAM, owing to a misinterpretation of the complex conjugate term in Eq. (9). We argue that only when $\Omega_2$ is considered as a vortex beam can the algebraic relation of topological charges in FWM be fully established (i.e., $l_s = l_p + l_{c1} - l_{c2}$). This omission has not been addressed in existing literature, yet it is essential for a complete characterization of OAM conservation in the FWM process. This relation is fundamental and may offer potential applications in quantum information encoding, quantum computing, and quantum communication.

Secondly, switching between matched vortex slow and fast light in the FWM process can be achieved by tuning the two strong control fields and the relative phase $\phi$ between the probe and signal fields. Such dispersion and absorption profiles have not been reported in previous studies, particularly regarding the sensitivity of the generated signal field's dispersive and absorptive characteristics to variations in the relative phase. Previous work [66], however, focused exclusively on the slow-light regime. In contrast, by employing a different system, we demonstrate that switching between slow and fast light can be realized by modulating both control fields, rather than varying a single one. This controllable modulation of light group velocity may offer potential utility in quantum information storage, all-optical signal processing, and ultrasensitive detection.

Our study demonstrates the potential to control and manipulate vortex light within a dual-Λ system, thereby enhancing our understanding of the interaction between vortex light and matter in this configuration.

**Disclosures**

The authors declare no conflicts of interest.

**Acknowledgements**

The authors acknowledge the National Natural Science Foundation of China (Grant Nos. 12474353, 12474354).

**References**

[1] R. W. Boyd and D. J. Gauthier, "Controlling the velocity of light pulses," Science 326(5956), 1074–1077 (2009).

[2] A. Kasapi, M. Jain, G. Y. Yin and S. E. Harris, "Electromagnetically Induced Transparency: Propagation Dynamics," Phys. Rev. Lett. 74, 2447–2450 (1995).

[3] P. L. Knight, B. Stoicheff, D. Walls (eds), "Highlights in quantum optics," Phil. Trans. R. Soc. Lond. A 355, 2215–2416 (1997).

[4] S. E. Harris "Electromagnetically Induced Transparency," Phys. Today 50(7), 36-42 (1997).

[5] K. J. Boller, A. Imamoğlu, and S. E. Harris, "Observation of electromagnetically induced transparency," Phys. Rev. Lett. 66(20), 2593–2596 (1991).

[6] M. Fleischhauer, A. Imamoğlu, and J. P. Marangos, "Electromagnetically induced transparency: Optics in coherent media," Rev. Mod. Phys. 77(2), 633–673 (2005).

[7] E. Paspalakis and P. Knight, "Electromagnetically induced transparency and controlled group velocity in a multilevel system," Phys. Rev. A 66(1), 015802 (2002).

[8] A. I. Lvovsky, B. C. Sanders, and W. Tittel, "Optical quantum memory," Nat.

Photonics 3(12), 706–714 (2009).

[9] I. Novikova, R. L. Walsworth, and Y. Xiao, "Electromagnetically induced transparency-based slow and stored light in warm atoms," Laser Photonics Rev. 6(3), 333–353 (2012).

[10] D. Han, H. Guo, Y. Bai, H. Sun, "Subluminal and superluminal propagation of light in an N-type medium," Phys. Lett. A 334, 243-248 (2005).

[11] J. Q. Shen and S. He, "Dimension-sensitive optical responses of electromagnetically induced transparency vapor in a waveguide," Phys. Rev. A 74(6), 063831 (2006).

[12] J.-Q. Zhang, S. Zhang, J.-H. Zou, L. Chen, W. Yang, Y. Li, and M. Feng, "Fast optical cooling of nanomechanical cantilever with the dynamical zeeman effect," Opt. Express 21(24), 29695–29710 (2013).

[13] K. McDonnell, L. F. Keary, and J. D. Pritchard, "Demonstration of a Quantum Gate Using Electromagnetically Induced Transparency," Phys. Rev. Lett. 129, 200501 (2022).

[14] H. Kang, G. Hernandez, and Y. Zhu, "Superluminal and slow light propagation in cold atoms," Phys. Rev. A 70, 01180(R) (2004).

[15] M. Sahrai, H. Tajalli, K. T. Kapale, and M. S. Zubairy, "Tunable phase control for subluminal to superluminal light propagation," Phys. Rev. A 70, 023813 (2004).

[16] K. Kim, H. S. Moon, C. Lee, S. K. Kim, and J. B. Kim, "Observation of arbitrary group velocities of light from superluminal to subluminal on a single atomic transition line," Phys. Rev. A 68, 013810 (2003).

[17] Y. Bai, H. Guo, D. Han, H. Sun, "Effects of spontaneously generated coherence on the group velocity in a V system," Phys. Lett. A 340, 342–346 (2005).

[18] Q. Liao, X. Xiao, W. Nie, N. Zhou, "Transparency and tunable slow-fast light in a hybrid cavity optomechanical system," Opt. Express 28(4), 5288 (2020).

[19] M. S. Bigelow, N. N. Lepeshkin, and R. W. Boyd, "Observation of Ultraslow Light Propagation in a Ruby Crystal at Room Temperature," Phys. Rev. Lett. 90, 113903 (2003).

[20] R. W. Boyd, "Slow and fast light: fundamentals and applications," J. Mod. Opt. 56, 1908–1915 (2009).

[21] S. E. Harris, J. E Field, and A. Kasapi, "Dispersive properties of electromagnetically induced transparency," Phys. Rev. A 46, R29 (1992).

[22] L. V. Hau, S. E. Harris, Z. Dutton, and C. H. Behroozi, "Light speed reduction to 17 metres per second in an ultracold atomic gas," Nature (London) 397, 594 (1999).

[23] M. M. Kash, V. A. Sautenkov, A. S. Zibrov, L. Hollberg, G. R. Welch, M. D. Lukin, Y. Rostovtsev, E. S. Fry, and M. O. Scully, "Ultraslow group velocity and enhanced nonlinear optical effects in a coherently driven hot atomic gas," Phys. Rev. Lett. 82, 5229 (1999).

[24] D. Budker, D.F. Kimball, S.M. Rochester, and V.V. Yashchuk, "Nonlinear Magneto-optics and Reduced Group Velocity of Light in Atomic Vapor with Slow Ground State Relaxation," Phys. Rev. Lett. 83, 1767 (1999).

[25] A.V. Turukhin, V.S. Sudarshanam, M.S. Shahriar, J. A. Musser, B. S. Ham, P. R.

Hemmer, "Observation of Ultraslow and Stored Light Pulses in a Solid," Phys. Rev. Lett. 88, 023602 (2002).

[26] M. Fleischhauer and M. D. Lukin, "Dark-state polaritons in electromagnetically induced transparency," Phys. Rev. Lett. 84, 5094 (2000).

[27] F. Beil, M. Buschbeck, G. Heinze, and T. Halfmann, "Light storage in a doped solid enhanced by feedback-controlled pulse shaping," Phys. Rev. A 81, 053801 (2010).

[28] K. Honda, D. Akamatsu, M. Arikawa, Y. Yokoi, K. Akiba, S. Nagatsuka, T. Tanimura, A. Furusawa, and M. Kozuma, "Storage and Retrieval of a Squeezed Vacuum," Phys. Rev. Lett. 100, 093601 (2008).

[29] U.-S. Kim, Y.-H. Kim, "Simultaneous Trapping of Two Optical Pulses in an Atomic Ensemble as Stationary Light Pulses," Phys. Rev. Lett. 129, 093601 (2022).

[30] S. Chu, S. Wong, "Linear Pulse Propagation in an Absorbing Medium," Phys. Rev. Lett. 48, 738 (1982).

[31] A.M. Steinberg, R.Y. Chiao, "Dispersionless, highly superluminal propagation in a medium with a gain doublet," Phys. Rev. A 49, 2071-2075 (1994).

[32] L.J. Wang, A. Kuzmich, A. Dogariu, "Gain-assisted superluminal light propagation," Nature (London) 406, 277-279 (2000).

[33] D. Ye, G. Zheng, J. Wang, Z. Wang, S. Qiao, J. Huangfu, L. Ran, "Negative group velocity in the absence of absorption resonance," Sci. Rep. 3, 1628 (2013).

[34] L. Brillouin, "Wave Propagation and Goup Velocity," (Academic Press, New York, 1960).

[35] N. Brunner, V. Scarani, M. Wegmuller, M. Legré, and N. Gisin, "Direct Measurement of Superluminal Group Velocity and Signal Velocity in an Optical Fiber," Phys. Rev. Lett. 93, 203902 (2004).

[36] D. Ye, Y. Salamin, J. Huangfu, S. Qiao, G. Zheng, and L. Ran, "Observation of wave packet distortion during a negative-group-velocity transmission," Sci. Rep. 5, 8100 (2015).

[37] M. D. Stenner, D. J. Gauthier, and M. A. Neifeld, "The speed of information in a 'fast-light' optical medium," Nature (London) 425, 695 (2003).

[38] M. D. Stenner, D. J. Gauthier, and M. A. Neifeld, "Fast Causal Information Transmission in a Medium With a Slow Group Velocity," Phys. Rev. Lett. 94, 053902 (2005).

[39] M. Tomita, H. Uesugi, P. Sultana, and T. Oishi, "Causal information velocity in fast and slow pulse propagation in an optical ring resonator," Phys. Rev. A 84, 043843 (2011).

[40] P. Bianucci, C. R. Fietz, J. W. Robertson, G. Shvets, and C.-K. Shih, "Observation of simultaneous fast and slow light," Phys. Rev. A 77, 053816 (2008).

[41] B. Macke and B. Ségard, "Simultaneous slow and fast light involving the Faraday effect," Phys. Rev. A 94, 043801 (2016).

[42] R. Pugatch, M. Shuker, O. Firstenberg, A. Ron, and N. Davidson, "Topological Stability of Stored Optical Vortices," Phys. Rev. Lett. 98, 203601 (2007).

[43] E. V. Barshak, C. N. Alexeyev, B. P. Lapin, and M. A. Yavorsky, "Twisted anisotropic fibers for robust orbital-angular-momentum-based information transmission," Phys. Rev. A 91, 033833 (2015).

[44] N. Radwell, T. W. Clark, B. Piccirillo, S. M. Barnett, and S. Franke-Arnold, "Spatially Dependent Electromagnetically Induced Transparency," Phys. Rev. Lett. 114, 123603 (2015).

[45] A. M. Akulshin, R. J. McLean, E. E. Mikhailov, and I. Novikova, "Distinguishing nonlinear processes in atomic media via orbital angular momentum transfer," Opt. Lett. 40(6), 1109 (2015).

[46] W. Zhen, X. L. Wang, J. Ding, and H.T. Wang, "Controlling the symmetry of the photonic spin Hall effect by an optical vortex pair," Phys. Rev. A 108, 023514 (2023).

[47] H.H. Wang, J. Wang, Z.H. Kang, L. Wang, J.Y. Gao, Y. Chen, X.J. Zhang, "Transfer of orbital angular momentum of light using electromagnetically induced transparency," Phys. Rev. A 100, 013822 (2019).

[48] Chanchal, G. P. Teja, C. Simon, and S. K. Goyal, "Storing vector-vortex states of light in an intra-atomic frequency-comb quantum memory," Phys. Rev. A 104, 043713 (2021).

[49] Y. Wen, I. Chremmos, Y. Chen, Y. Zhang, and S. Yu, "Arbitrary Multiplication and Division of the Orbital Angular Momentum of Light," Phys. Rev. Lett. 124, 213901 (2020).

[50] F. Meng, X.G. Wei, Y.J. Qu, Y. Chen, X.J. Zhang, Z.H. Kang, L.Wang, H.H. Wang, J.Y. Gao, "Arithmetic operation of orbital angular momentum of light via slow-light four-wave mixing," J. Lumin. 242, 118551 (2022).

[51] S. Li, X. Pan,1, Y. Ren, H. Liu, S. Yu, and J. Jing, "Deterministic Generation of Orbital-Angular-Momentum Multiplexed Tripartite Entanglement," Phys. Rev. Lett. 124, 083605 (2020).

[52] L. Wang, X. Zhang, A. Li, Z. Kang, H. Wang, J. Gao, "Controlled-not gate with orbital angular momentum in a rare-earth-ion-doped solid," J. Lumin. 228, 117628 (2020).

[53] H. R. Hamedi, J. Ruseckas, E. Paspalakis, and G. Juzeliunas, "Transfer of optical vortices in coherently prepared media," Phys. Rev. A 99, 033812 (2019).

[54] M. Mahdavi, Z. Amini Sabegh, M. Mohammadi, M. Mahmoudi, and H. R. Hamedi, "Manipulation and exchange of light with orbital angular momentum in quantum-dot molecules," Phys. Rev. A 101, 063811 (2020).

[55] S. H. Asadpour, Ziauddin, M. Abbas, and H. R. Hamedi, "Exchange of orbital angular momentum of light via noise-induced coherence," Phys. Rev. A 105, 033709 (2022).

[56] Z. Wang, Y. Zhang, E. Paspalakis, and B. Yu, "Efficient spatiotemporal-vortex four-wave mixing in a semiconductor nanostructure," Phys. Rev. A 102, 063509 (2020).

[57] M. Mahdavi and Z. A. Sabegh, "Orbital angular momentum transfer in molecular magnets," Phys. Rev. B 104, 094432 (2021).

[58] C. Meng, T. Shui, and W. X. Yang, "Coherent transfer of optical vortices via

backward four-wave mixing in a double-Λ atomic system," Phys. Rev. A 107, 053712 (2023).

[59] R. Kumar, D. Manchaiah, M. Ahmad and R.K. Easwaran, "Effect of relaxation on the transfer of orbital angular momentum via four-wave mixing process in the four-level double lambda atomic system," New J. Phys. 26, 053045 (2024).

[60] M. Padgett and R. Bowman, "Tweezers with a twist," Nat. Photonics 5(6), 343-348 (2011).

[61] P. Senthilkumaran, "Optical phase singularities in detection of laser beam collimation," Appl. Opt. 42(31), 6314-6320 (2003).

[62] K. Masuda, S. Nakano, D. Barada, M. Kumakura, K. Miyamoto, and T. Omatsu, "Azo-polymer film twisted to form a helical surface relief by illumination with a circularly polarized Gaussian beam," Opt. Express 25, 12499-12507 (2017).

[63] P. S. Tan, X. C. Yuan, G. H. Yuan, Q. Wang, "High-resolution wide-field standing-wave surface plasmon resonance fluorescence microscopy with optical vortices," Appl. Phys. Lett. 97(24), 241109 (2010).

[64] M. Padgett, J. Courtial, and L. Allen, "Light's orbital angular momentum," Phys. Today 57(5), 35 (2004).

[65] M. Babiker, D. L. Andrews, and V. E. Lembessis, "Atoms in complex twisted light," J. Opt. 21, 013001 (2019).

[66] H. R. Hamedi, I. A. Yu, and E. Paspalakis, "Matched optical vortices of slow light using a tripod coherently prepared scheme," Phys. Rev. A 108, 053719 (2023).

[67] L. Allen, M. J. Padgett, and M. Babiker, "IV The Orbital Angular Momentum of Light," Prog. Opt. 39, 291 (1999).